\documentstyle[12pt]{article}

\input{tcilatex}

\begin{document}

\title{Compositeness and the asymmetries of leptons at the $Z^0$ peak}
\author{R. D\'{\i}az Sanchez.$^{1}$, R. Mart\'{\i}nez$^{1}$ and J.-Alexis
Rodr\'{\i}guez$^{1,2}$ \\
1. Depto. de F\'{\i}sica, Universidad Nacional, Bogot\'a, Colombia \\
2. Centro Internacional de F\'{\i}sica, Bogot\'a, Colombia}
\date{}
\maketitle

\begin{abstract}
We study the effects on the leptonic asymmetries $A_{FB}$ and $A_{LR}$
coming from a model of compositeness. We consider the effects coming from
the self-energies and the vertex correcction to $Zl^{+}l^{-}$. Thus we use
the Altarelli parametrization of the oblique corrections. We get the
asymptotic limits of these corrections in terms of the parameters $(m^{\ast
},\ \Lambda ,\ f$ $,f^{\prime })$ and we get bounds for the quotient $%
m^{\ast }/\Lambda $ with different values of $\ (f$ $,f^{\prime })$. We
conclude that both asymmetries produce bounds for such quotient when $%
f^{\prime }$ overweight to $f$, and this fact is related with the breakdown
of the custodial symmetry.
\end{abstract}

\section{Introduction}

Compositeness \cite{pati} is regarded as a solution to many problems which
are unsolved in the Standard Model, like the fermion mass spectrum, the
generation of masses and the large number of parameters. The idea is that
some underlying structure can explain the quark and lepton family pattern.
At this stage, we shall search for new excited fermions ($\psi _F^{*}$), and
for new interactions originated from the constituents.

A signal for a composite structure of fermions could be the production of
excited leptons or quarks, either pairwise, due to their normal gauge
couplings, or singly, due to radiative transitions between normal and
excited fermions \cite{baur} \cite{hawigara}. For instance, excited fermions
could be produced in $Z$ decays; in fact, there are strong limits coming
from the $Z^0$ partial widths. Mass regions up to $45$ GeV for $u^*$ and $%
d^* $ as well as $35-40$ GeV for $l^*$ and $\nu^*$ have been excluded \cite
{lep}. However, the $l^*$ limits can be pushed to $46$ GeV by searching the
decay mode $Z \to l^* \bar l^* \to l^+ l^- \gamma \gamma$ \cite{lep}.

Moreover, recently the possibility has been studied that the significant
excess from QCD predictions found by the collider detector at Fermilab (CDF)
in the inclusive jet cross section for transverse energies $E_T \geq 200$
GeV could be explained by the production of excited bosons or excited quarks
in the mass region of $1600$ GeV and $500$ GeV\cite{catani} \cite{terazawa},
respectively. Other studies have already been done in order to get the low
energy scale $\Lambda$ for compositeness by precise comparison between
currently available experimental data and calculations in the composite
model of quarks \cite{akama}.

On the other hand, radiative corrections are the appropriate place to
indirectly look for new physics at low energies \cite{akama} \cite{martinez}%
, because when heavy fermions are coupled to longitudinal components of the
gauge bosons there is no decoupling of the high energy physics from the low
energy physics. In ref. \cite{martinez}, radiative corrections at the $Z$
scale have been considered in order to bound substructure. Our purpose in
this paper is to find bounds for the masses of possible excited leptons by
using the radiative corrections involved in the forward-backward and
left-right asymmetries for the lepton average.

\section{The model}

In general, the normal fermions ($\psi_f$) and excited fermions ($\psi_F^*$)
at low energies can be described by an effective Lagrangian like \cite
{hawigara} 
\begin{equation}
{\cal L}_{eff}= \sum_{i=\gamma,Z,W^\pm} v_i \bar \psi_F^* \gamma^\mu V_\mu^i
\psi_F^* +\frac e \Lambda \bar \psi^*_F \sigma^{\mu \nu} (c_{iFf}-d_{iFf}
\gamma_5) \psi_f V^i_{\mu \nu} +h.c. \; \; ,
\end{equation}
where $V^i_{\mu \nu}$ are the field strengths for the gauge bosons and $%
\Lambda$ denotes the compositeness scale. In particular, we choose a
specific model which assumes that excited leptons form a weak isodoublet, 
\begin{equation}
\bar L=(\bar \nu^*, \bar l^*)
\end{equation}
which couples to the ordinary left handed lepton doublet by the $SU(2)_L
\otimes U(1)_Y$ invariant interaction Lagrangian 
\begin{equation}
{\cal L}=\frac {gf}{\Lambda} \bar L \sigma^{\mu \nu} \frac {\tau}{2} l_L
W_{\mu \nu} +\frac {g^{\prime}f^{\prime}}{\Lambda} \bar L \sigma^{\mu \nu}
\frac {Y}{2} l_L B_{\mu \nu}+h.c. \; \; .
\end{equation}
Where $g$ and $g^{\prime}$ are the $SU(2)$ and $U(1)$ coupling constants
respectively, $\tau$ denotes the Pauli matrices and $Y$ is the hypercharge.

As for the coefficients $c_{iFf}$ and $d_{iFf}$, in general, $g-2$
measurements imply $|c|=|d|$, and also, the absence of dipole moments for
the electron and the muon requires $c$ and $d$ to be relatively real \cite
{renard}. Thus, we have the following relations:

\begin{eqnarray}
c_{\gamma l^*l}&=&-\frac 14 (f+f^{\prime}) \;\; ,  \nonumber \\
c_{\gamma \nu^* \nu}&=&\frac 14 (f-f^{\prime})\; \; ,  \nonumber \\
c_{Z l^* l}&=&-\frac 14 (f \cot \theta_W +f^{\prime}\tan \theta_W) \; \; , 
\nonumber \\
c_{Z \nu^* \nu}&=&\frac 14 (f \cot \theta_W +f^{\prime}\tan \theta_W) \; \; ,
\nonumber \\
c_{W^\pm \nu^* l}&=&\frac {f}{2 \sqrt{2} \sin \theta_W} \; \; .
\end{eqnarray}

Now, since the excited fermions are doublets, their couplings to photon, $Z$
and $W$ bosons are defined by the following renormalizable Lagrangian 
\begin{equation}
{\cal L}=\bar{L}\gamma _\mu (g\frac \tau 2\cdot W^\mu +\frac{g^{\prime }}%
2YB^\mu )L\;\;,
\end{equation}
where the coefficients $v_i$ from eq.(1) satisfy 
\begin{eqnarray}
v_{\gamma l^{*}l} &=&-e\;\;,  \nonumber \\
v_{\gamma \nu ^{*}\nu } &=&0\;\;,  \nonumber \\
v_{Zl^{*}l} &=&-\frac e{2s_Wc_W}(1-2s_W^2)\;\;,  \nonumber \\
v_{Z\nu ^{*}\nu } &=&\frac e{2s_Wc_W}\;\;,  \nonumber \\
v_{W^{+}\nu ^{*}l} &=&\frac e{\sqrt{2}s_W}\;\;.
\end{eqnarray}

\section{Self-energy contributions}

Once we have the framework setup (eqs. (1)-(6)), we can calculate the
contribution to the self-energies of the gauge bosons by using dimensional
regularization, where the pole $d=4$ is identified with $\ln (\Lambda
^{2}/m_{Z}^{2})$ \cite{zepenfeld}. In order to simplify the analysis we
assume the same mass for the excited leptons, $m_{\nu }^{\ast }=m_{l}^{\ast
}=m^{\ast }$; this is justified by the fact that we are considering the
excited fermion in a linear vectorial representation of the $SU(2)\times
U(1) $ gauge group. The states with equal mass for $m_{\nu }^{\ast }$ and $%
m_{l}^{\ast }$ do not break the symmetry.

In general the self-energies due to the lagrangians of dimension 5 and 4
are: 
\begin{eqnarray}
\Sigma _{V_{i}V_{j}}^{(5)}(q^{2}) &=&-\frac{c_{V_{i}}c_{V_{j}}\alpha }{9\pi
\Lambda ^{2}}F^{(5)}(m^{\ast },q^{2})  \nonumber \\
\Sigma _{V_{i}V_{j}}^{(4)}(q^{2}) &=&-\frac{v_{i}v_{j}\alpha }{12\pi ^{2}}%
F^{(4)}(m^{\ast },q^{2})
\end{eqnarray}
where the functions $F(m^{\ast },q^{2})$ can be written as \cite{martinez} 
\begin{eqnarray}
F^{(5)}(m^{\ast },q^{2}) &=&6m^{\ast 4}[1-(\frac{m^{\ast 2}-q^{2}}{q^{2}}%
)\ln (\frac{m^{\ast 2}}{m^{\ast 2}-q^{2}})]  \nonumber \\
&+&3m^{\ast 2}q^{2}[1+(\frac{m^{\ast 2}-q^{2}}{q^{2}})\ln (\frac{m^{\ast 2}}{%
m^{\ast 2}-q^{2}})]  \nonumber \\
&-&q^{4}[8+3\ln (\frac{\Lambda ^{2}}{m^{\ast 2}})-3(\frac{m^{\ast 2}-q^{2}}{%
q^{2}})\ln (\frac{m^{\ast 2}}{m^{\ast 2}-q^{2}})\;\;,  \nonumber \\
F^{(4)}(m^{\ast },q^{2}) &=&q^{2}\ln \frac{\Lambda ^{2}}{m^{\ast 2}}%
+4m^{\ast 2}+\frac{5}{3}q^{2}  \nonumber \\
&-&2(2m^{\ast 2}+q^{2})\sqrt{\frac{4m^{\ast 2}}{q^{2}}-1}\arctan (\frac{1}{%
\sqrt{\frac{4m^{\ast 2}}{q^{2}}-1}})\;\;.
\end{eqnarray}
We have considered the ordinary lepton masses negligible. Therefore, we find
the vacuum polarization tensors for the gauge bosons: 
\begin{eqnarray}
-\Pi _{Z}(m_{Z}^{2}) &=&\frac{\alpha (1-2s_{W}^{2}+2s_{W}^{4})}{6\pi
m_{Z}^{2}s_{W}^{2}c_{W}^{2}}F^{(4)}(m^{\ast },m_{Z}^{2})  \nonumber \\
&&+\frac{\alpha }{72\pi m_{Z}^{2}\Lambda ^{2}}(f\cot \theta _{W}+f^{\prime
}\tan \theta _{w})^{2}F^{(5)}(m^{\ast },m_{Z}^{2})\text{ \ ,}  \nonumber \\
-\Pi _{\gamma Z}(m_{Z}^{2}) &=&\frac{\alpha (1-2s_{W}^{2})}{6\pi
s_{W}c_{W}m_{Z}^{2}}F^{(4)}(m^{\ast },m_{Z}^{2})  \nonumber \\
&&+\frac{\alpha f}{72\pi \Lambda ^{2}m_{Z}^{2}}(f\cot \theta _{W}+f^{\prime
}\tan \theta _{w})F^{(5)}(m^{\ast },m_{Z}^{2})\text{ \ \ ,}  \nonumber \\
-\Pi _{\gamma }(m_{Z}^{2}) &=&\frac{\alpha }{3\pi m_{Z}^{2}}F^{(4)}(m^{\ast
},m_{Z}^{2})  \nonumber \\
&&+\frac{\alpha }{72\pi \Lambda ^{2}m_{Z}^{2}}(f^{2}+f^{\prime
2})F^{(5)}(m^{\ast },m_{Z}^{2})\text{ \ \ ,}
\end{eqnarray}
where $\Pi _{W}(0)=0$ and $\Sigma _{Z}^{\prime }(m_{Z}^{2})$ is obtained
directly from $\Pi _{Z}(m_{Z}^{2})$. The asymptotic limits $\left( m^{\ast
},\Lambda \right) \rightarrow \infty \;$in the functions $F(m^{\ast
},q^{2})\;$are

\begin{eqnarray}
\frac{F^{\left( 5\right) }(m^{\ast },q^{2})}{\Lambda ^{2}} &=&9\left( \frac{%
m^{\ast }}{\Lambda }\right) ^{2}q^{2}\text{ \ ,}  \label{asyn} \\
F^{\left( 4\right) }(m^{\ast },q^{2}) &=&q^{2}\ln \left( \frac{\Lambda }{%
m^{\ast }}\right) ^{2}\text{ \ .}  \nonumber
\end{eqnarray}
In such limit, $\Sigma _{V_{i}V_{j}}^{(5)}(q^{2})$ approaches a constant
value and the physics described by (3) is non-decoupled.

\section{The asymmetries and the new physics}

On the other hand, the forward-backward asymmetry $A_{FB}$ and the
left-right asymmetry $A_{LR}$ for leptons in the $Z$ decays are given by 
\begin{eqnarray}
A_{FB}^{l} &=&\frac{g_{V}^{e}g_{A}^{e}}{g_{V}^{e2}+g_{A}^{e2}}\frac{%
g_{V}^{l}g_{A}^{l}}{g_{V}^{l2}+g_{A}^{l2}}\;\;,  \nonumber \\
A_{LR} &=&\frac{2g_{V}^{e}g_{A}^{e}}{g_{V}^{e2}+g_{A}^{e2}}\;\;.
\end{eqnarray}
where a $Zll$ amplitude is defined as $-ig\gamma _{\mu
}(g_{V}^{l}-g_{A}^{l}\gamma _{5})/4c_{W}$ and the superscripts $e,l$ denote
electron, lepton respectively. In our case, new physics is included in the
constants $g_{V,A}^{l}$ as 
\begin{equation}
g_{V,A}^{l}=g_{V,A}^{SM}+\delta g_{V,A}^{l}
\end{equation}
Where the superindex SM denotes the SM coupling with the contribution at one
loop level of the top quark and Higgs scalar boson. In this way, $\delta
g_{V,A}^{l}$ only contains the new physics contribution. With this
prescription we find that $A_{FB}^{l}$ and $A_{LR}$ can be written as 
\begin{eqnarray}
A_{FB}^{l} &=&A_{FB}^{l,\;SM}(1+2\delta ^{NP})\;\;,  \nonumber \\
A_{LR} &=&A_{LR}^{SM}(1+\delta ^{NP})\;\;,
\end{eqnarray}
where $\delta ^{NP}$ has the contribution of new physics and 
\begin{equation}
\delta ^{NP}=\frac{\delta g_{A}^{l}}{g_{A}^{SM}}+\frac{\delta g_{V}^{l}}{%
g_{V}^{SM}}-2\frac{g_{V}^{SM}\delta g_{V}^{l}+g_{A}^{SM}\delta g_{A}^{l}}{%
(g_{V}^{SM})^{2}+(g_{A}^{SM})^{2}}\;\;.
\end{equation}

Additionally, these couplings $\delta g_{V,A}^{l}$ can be expressed in terms
of the $\epsilon _{i}$ parameters \cite{altarelli}, 
\begin{eqnarray}
\delta g_{V} &=&=\frac{2s_{W}^{2}}{c_{W}^{2}-s_{W}^{2}}\epsilon _{3}-(\frac{1%
}{4}+\frac{s_{W}^{2}}{c_{W}^{2}-s_{W}^{2}})\epsilon _{1}-\epsilon _{l}/2\;\;,
\nonumber \\
\delta g_{A} &=&-\frac{1}{4}\epsilon _{1}-\epsilon _{l}/2\;\;.
\end{eqnarray}
Here, $\epsilon _{l}$ is the $Zl^{+}l^{-}$ vertex correction given in \cite
{martinez}. The other two parameters $\epsilon _{1}$ and $\epsilon _{3}$ can
be written as a function of self-energies as given by eq. (9) \cite
{altarelli}.

We consider the asymmetries for electron and leptonic average. The SM
predicted values are \cite{hawi} 
\begin{eqnarray}
A_{FB}^{SM} &=&0.0168\;\;,  \nonumber \\
A_{LR}^{SM} &=&0.1485\;\;
\end{eqnarray}
for $\alpha =1/128.75$, $m_{W}=80.35$, $M_{H}=100$ GeV and $m_{t}=175$ GeV.
And the experimental values are \cite{data} : 
\begin{eqnarray}
A_{FB}^{l} &=&0.0174\pm 0.0010\;\;,  \nonumber \\
A_{LR} &=&0.1542\pm 0.0037\;\;.
\end{eqnarray}

Notice that the standard model value $A_{LR}\;$goes 1.54$\sigma \;$out from
experimental data. Figure 1 shows the $A_{FB}$ as a function of the scale $%
\Lambda $ for values of $m^{\ast }=100$ GeV (a) and $m^{\ast }=500$ GeV (b).
Different values of the parameters $f$ and $f^{\prime }$ have been
considered. Figure 2 shows $A_{LR}$ as a function of the scale $\Lambda $
for the excited lepton mass of $m^{\ast }=500$ GeV (a) and $m^{\ast }=1000$
GeV (b). In figure 3 we have plotted $A_{FB}$ of the electron versus $%
m^{\ast }/\Lambda $, as $\Lambda $ is the energy scale for new physics, $%
m^{\ast }$ must be less than $\Lambda $, and we have $m^{\ast }/\Lambda \;$%
between $0\;$and $1$. In figure 4 we show the left-right asymmetry versus
the quotient $m^{\ast }/\Lambda $ for different values of $f\;$and $%
f^{\prime }\;$couple constants.\ 

We can see that both forward-backward and left-right asymmetries produce
bounds for such quotient when $f^{\prime }\;$overweight considerably to $f,\;
$it means that custodial symmetry is strongly broken and therefore radiative
contributions become higher$.$

\section{Conclusion}

In conclusion, we have evaluated the contribution of excited lepton states,
up to one-loop level, to the oblique parameters. We have included this
contribution into the leptonic asymmetries $A_{FB}\;$and $A_{LR}$ , and we
have compared our results with the precise data on the electroweak
observables obtained by LEP and LSD Collaborations \cite{data} \cite{hawi}.
Therefore, we can extract bounds on the compositeness scale and the excited
lepton mass in the context of the phenomenological model under
consideration. Further, in equation (3) the term proportional to $B_{\mu \nu
}$ breakdowns the custodial symmetry and it is proportional to the constant $%
f^{\prime }$. Otherwise the new physics is allowed in the experimental
region ($1\sigma )$, which is satisfied when custodial symmetry is strongly
broken. \ In order to get this requirement we need that $f^{\prime }$
becomes bigger than $f$. With this prescription we obtain bounds on \ the
rate $m^{\ast }/\Lambda .$

\section{Acknowlegments}

We thank F. Larios for reading the manuscript and useful comments. We
acknowledge the financial support from COLCIENCIAS (Colombia) and the
hospitality of CINVESTAV (Mexico) during the realization of this work. R. D.
express his acknowledgements to the Fundaci\'{o}n MAZDA Para el Arte y la
Ciencia for his fellowship.

\newpage

\begin{center}
Figure Captions
\end{center}

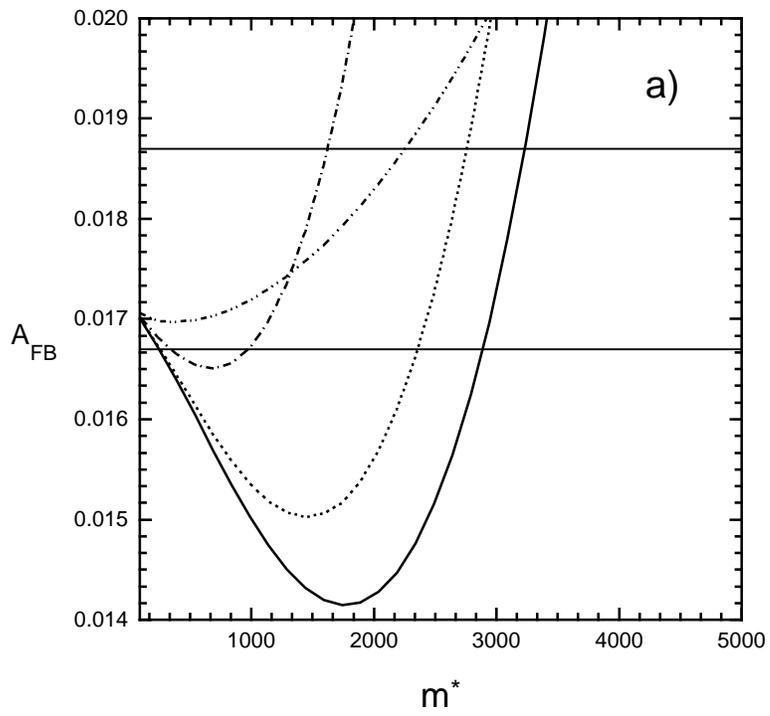
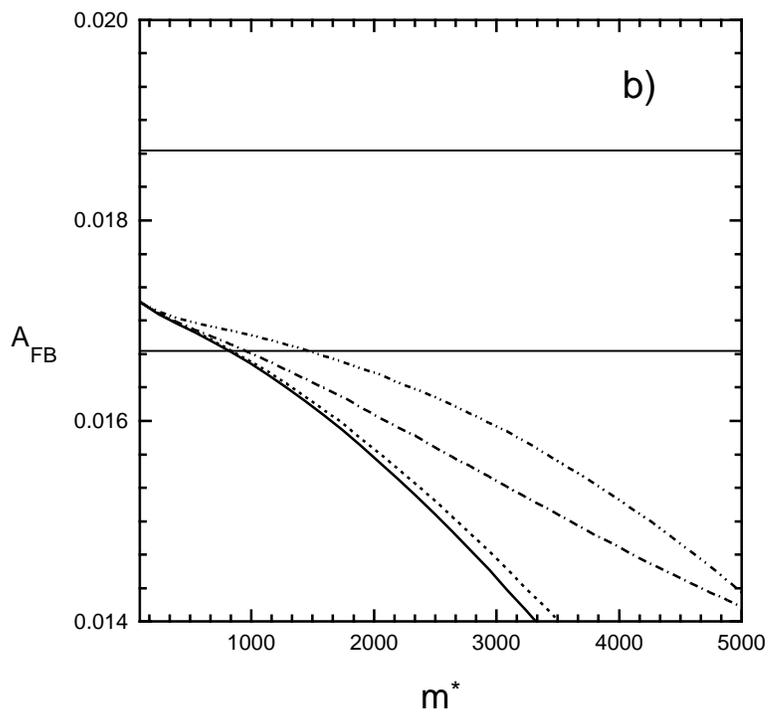
\noindent Figure 1. $A_{FB}$ as a function of the scale $\Lambda $ for
values of the excited lepton mass of $m^{\ast }=100$GeV (a) and $m^{\ast
}=500$GeV (b). From the upper solid line we use the values ( $f=f^{\prime
}=1 $ ), ( $f=-f^{\prime }=1$ ), ( $f=0$, $f^{\prime }=1)$ up to ( $f=1$ ,$%
f^{\prime }=0)$ the lower dot-dot-solid line. The horizontal solid lines are
the experimental limits.

\vspace{1cm}

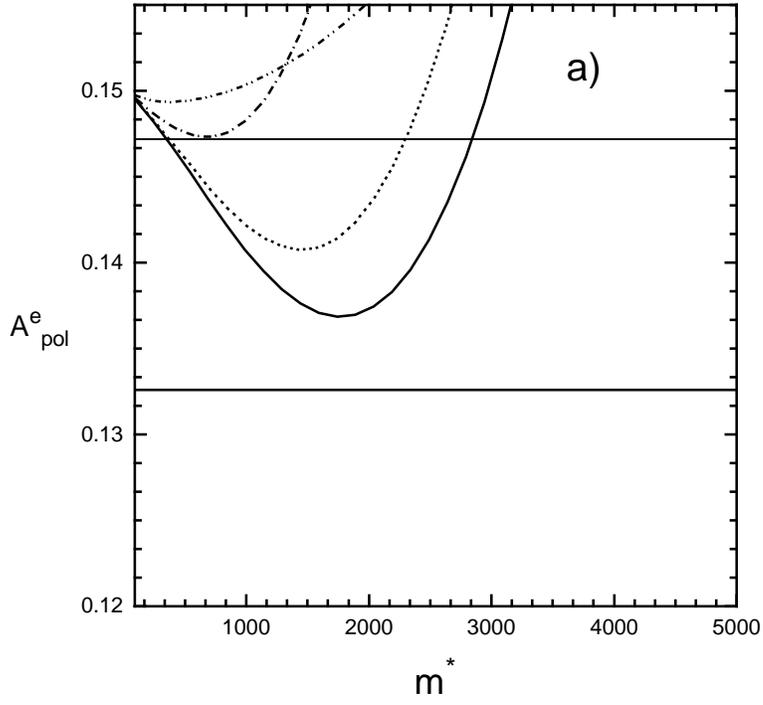
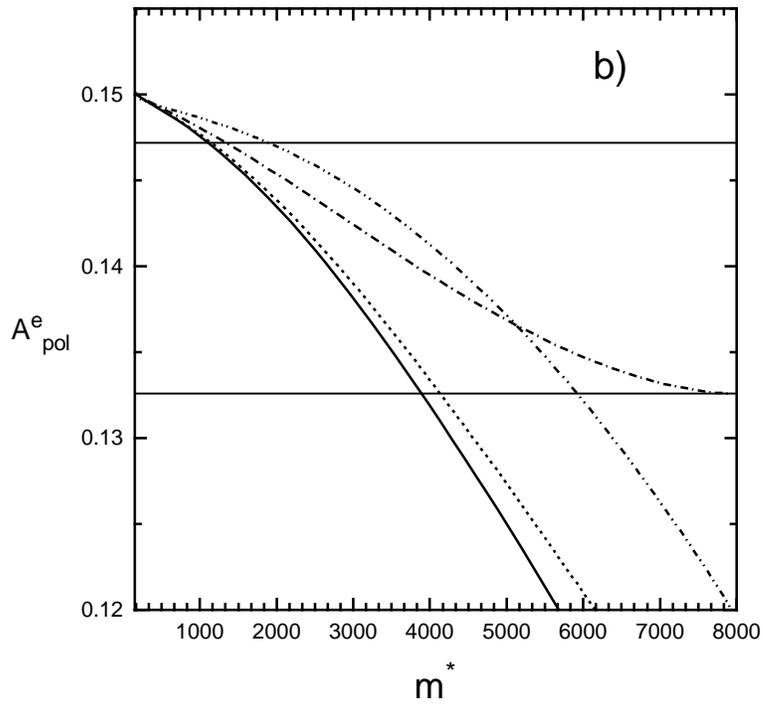
\noindent Figure 2. As figure 1 for $A_{LR}$.

\vspace{1cm}

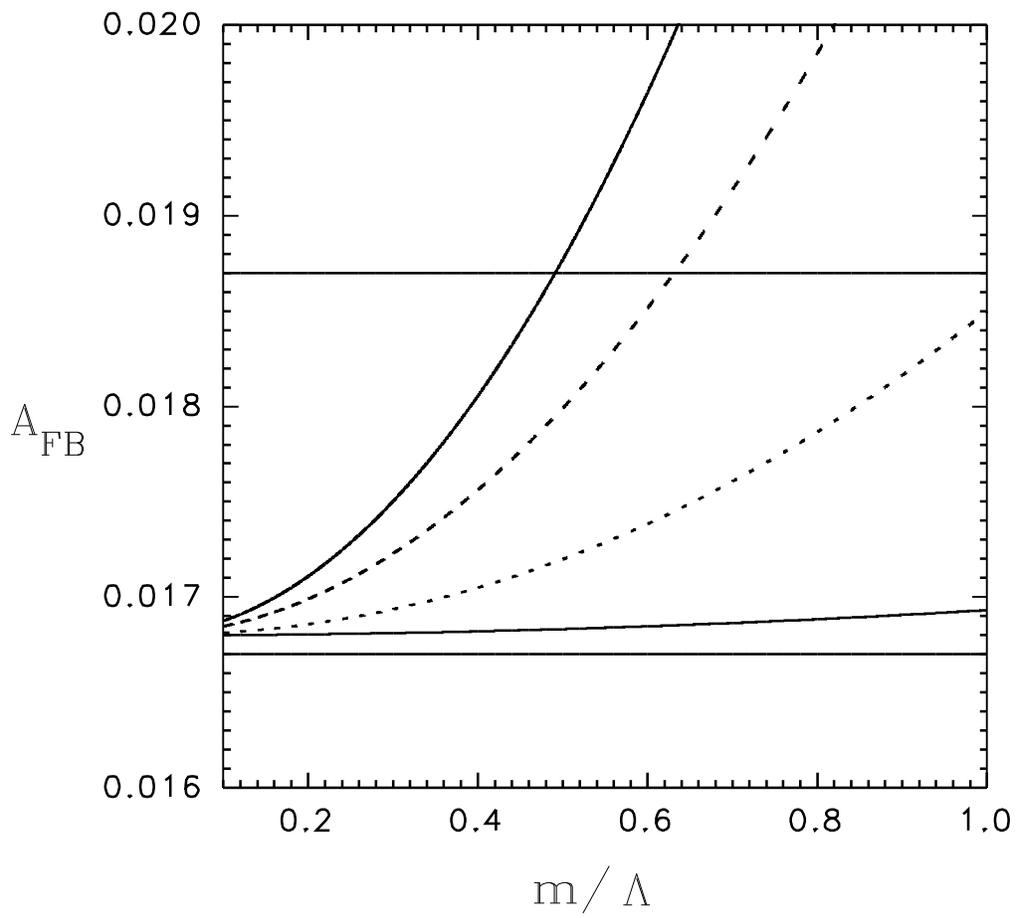
\noindent Figure 3. $A_{FB}$ as a function of the ratio $m^{\ast }/\Lambda $
. From the upper solid line we use the values ($f=1$ , $f^{\prime }=7)$ , ( $%
f=0$ , $f^{\prime }=6$ ), ( $f=-1$, $f^{\prime }=3)$ up to ( $f=1$ , $%
f^{\prime }=-1$ ) the lower dashed line. The horizontal solid lines are the
experimental limits.

\vspace{1cm}

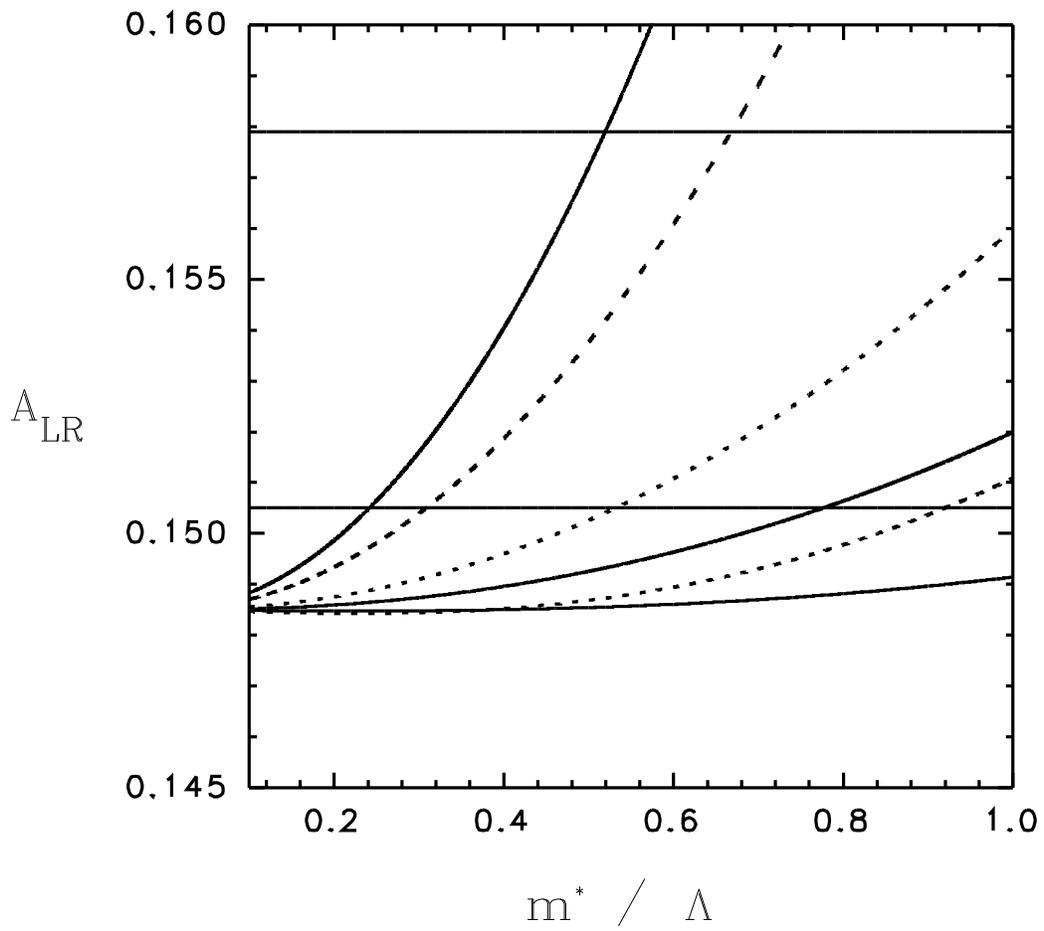
\noindent Figure 4. $A_{LR}$ as a function of the ratio $m^{\ast }/\Lambda $
. From the upper solid line we use ($f=1$ , $f^{\prime }=7)$ , ( $f=0$ , $%
f^{\prime }=6$ ) , \ ( $f=-1$, $f^{\prime }=3$ ), ( $f=-2$, $f^{\prime }=2$
), ( $f=1$, $f^{\prime }=2$ ), \ up to $f=1$ ,$f^{\prime }=-1$ (lower dashed
line). The horizontal solid lines are the experimental limits.

\end{document}